\documentclass{ws-ijmpa}
\usepackage{times}

\newcommand{\Xmax}{X_{\max}}
\newcommand{\Chis}{\chi^{2}}
\newcommand{\Nexp}{N_{\exp}}
\newcommand{\Nteor}{N_{\mathrm{theor}}}
\newcommand{\Ai}{A_{\mathrm{i}}}
\newcommand{\E}{E_{0}}
\newcommand{\Ebar}{\overline{E}_{0}}

\begin{document}

\markboth{S.~P.~Knurenko et al.}
         {Fluctuations of $\Xmax$ and Primary Particle Mass Composition
          in the Range of Energy $5\times10^{17}-3\times10^{19}$~eV}

\catchline{}{}{}{}{}

\title{FLUCTUATIONS OF $\Xmax$ AND PRIMARY PARTICLE MASS COMPOSITION IN THE
       RANGE OF ENERGY $5\times10^{17}-3\times10^{19}$~EV BY YAKUTSK DATA}

\author{S.~P.~KNURENKO$^*$, A.~A.~IVANOV, V.~A.~KOLOSOV, Z.~E.~PETROV, \\
        I.~YE.~SLEPTSOV and G.~G.~STRUCHKOV}
\address{Yu.~G.~Shafer Institute of Cosmophysical Research and Aeronomy
         Lenin Avenue 31, Yakutsk 677891, Russia \\
         ${}^*$s.p.knurenko@ikfia.ysn.ru
       }

\maketitle

\pub{Received (24 October 2004)}{Revised (5 July 2005)}

\begin{abstract}
The experimental distributions of $\Xmax$ obtained with the Yakutsk EAS array
at fixed energies of $5 \times10^{17}$, $1\times10^{18}$ and $5\times10^{18}$~eV
are analysed. A recent version of the QGSJET model is used as a tool of our
analysis. In the framework of this model, the most adequate mass composition
of primary particles satisfying the experimental data on $\Xmax$ is selected.

\end{abstract}

\section*{}

A large body of data on charged particles, muons and Cherenkov radiation of EAS
in the energy range $10^{17}-10^{19}$~eV has been accumulated with the Yakutsk
EAS array over a thirty-year period of its exposure. During this period,
$\sim 10^{6}$ showers with $\E\ge10^{17}$~eV have been registered.
After 1993, the number of detectors for the charged particles and EAS Cherenkov
light has been increased. The accuracy for determination of arrival direction
(zenith and azimuth angles) of EAS and the EAS axis location have been
considerably improved.
This allowed us to estimate more precisely the EAS characteristics, including
$\Xmax$, the atmospheric depth at which the shower reaches its maximum.

A large number of Cherenkov detectors operating in the individual EAS events
as well as the use of a new version of the QGSJET model\cite{QGSJET} allowed
us to perform quantitative estimations of the mass composition of Primary Cosmic
Rays (PCR). For this aim we have compared the experimental data (see Fig.~1)
and theoretical predictions obtained with the QGSJET model for different primary
nuclei by applying the conventional $\Chis$ criterion. The value of $\Chis$ is
determined according to the equality
\begin{equation}
\Chis\left(\Xmax\right)=\frac{\sum\left[\Nexp\left(\Xmax\right)
-\Nteor\left(\Xmax\right)\right]^{2}}{\Nteor\left(\Xmax\right)},
\label{eq1}
\end{equation}
in which
\begin{equation}
\Nteor\left(\Xmax\right) = \sum_i~P\left(\Ai\right)
                 \cdot \Nteor\left(\Xmax,\Ai\right).
\label{Ntheor}
\end{equation}

\begin{figure}[h!]
\centering
\includegraphics[width=0.95\linewidth]%
 {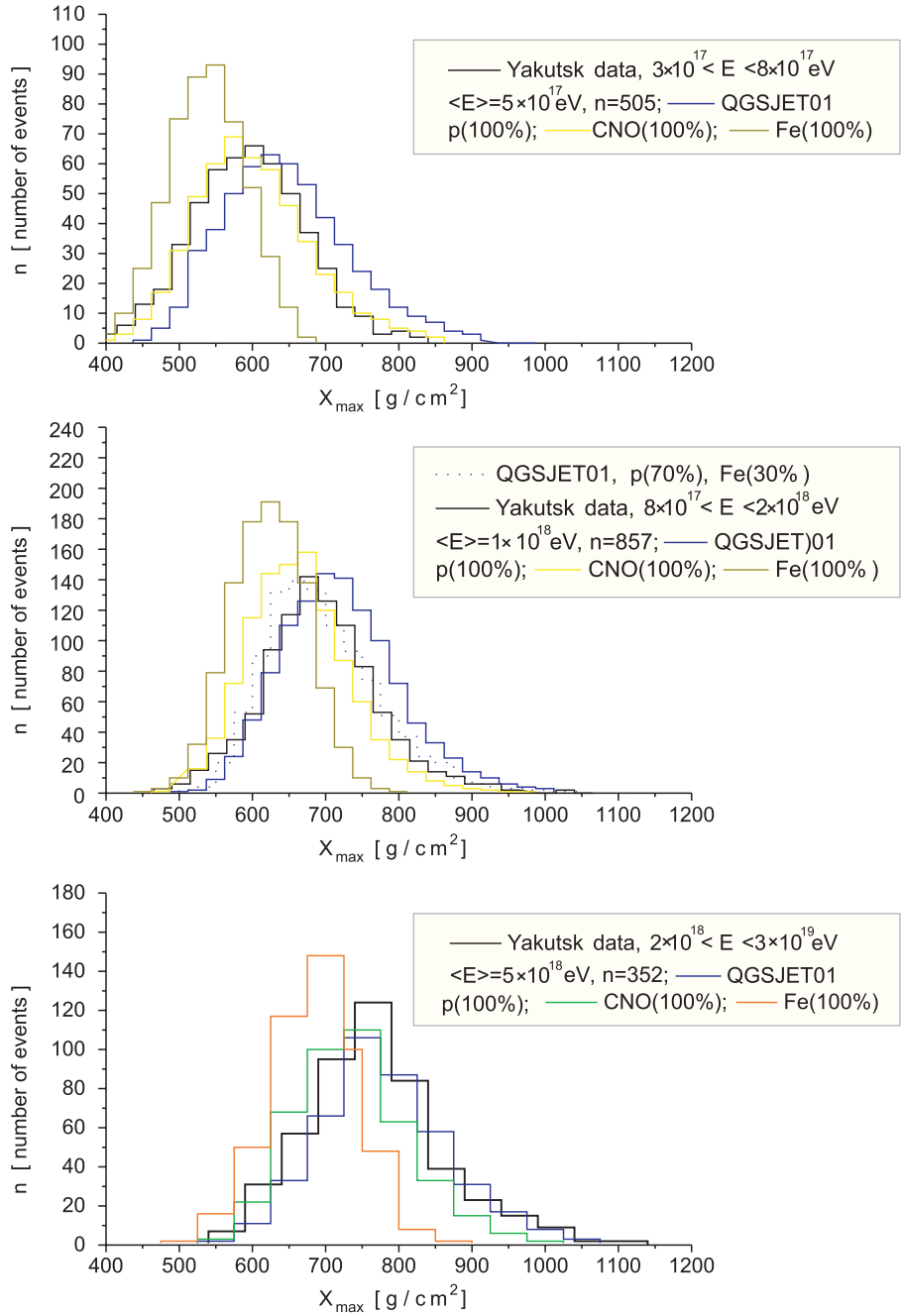}
\vspace*{8pt}
\caption{Distributions of depths of the shower maximum for different values
         of energy.}
\end{figure}

In Eqs.~(\ref{eq1}) and (\ref{Ntheor}), $\Nexp(\Xmax)$ is the experimental
number of showers in the interval from $\Xmax$ to $\Xmax+\Delta\Xmax$,
$\Nteor(\Xmax,\Ai)$ is the corresponding number of showers calculated under
the assumption that the mass number of the nucleus is equal to $\Ai$, and
$P(\Ai)$ is the probability that the shower with energy $\E$ is formed
by a primary particle with the mass number $\Ai$.

Our analysis involves five standard groups of primary nuclei:
p, $\alpha$, M, H and Fe.
The main results of the analysis are summarized in Tables~1 and 2.
\begin{table}[h!]
\tbl{$\Chis$ values (${\rm C.L.}=95\%$) for three average energies
     and five standard nuclear groups.}
{\begin{tabular}{@{}cccccc@{}}
\toprule
$\Ebar$ (eV)     &   p    & $\alpha$ &   M    &  H     &  Fe    \\
\colrule
$5\times10^{17}$ & $1.53$ &  $1.46$  & $1.24$ & $1.61$ & $2.61$ \\
$1\times10^{18}$ & $1.36$ &  $1.22$  & $1.09$ & $1.77$ & $2.76$ \\
$5\times10^{18}$ & $0.75$ &  $1.05$  & $1.14$ & $2.36$ & $3.25$ \\
\botrule
\end{tabular}}
\end{table}

The shape of the experimental $\Xmax$ distribution at the optimal value of
$\Chis$ with the confidence level of $0.95$ (see Table~1)
does not contradict to the mass compositions listed in Table~2 for
the same average energies and nuclear groups.
\begin{table}[h!]\label{tab2}
\tbl{Mass composition of the PCR for five nuclei components.}
{\begin{tabular}{@{}cccccc@{}}
\toprule
$\Ebar$ (eV)     &  
          p      &   $\alpha$    &     M         &      H        &       Fe      \\
\colrule
$5\times10^{17}$ &
  $(39\pm11)$\%  & $(31\pm13)$\% & $(18\pm10)$\% & $( 7\pm 6)$\% & $( 5\pm 4)$\% \\
$1\times10^{18}$ &
  $(41\pm 8)$\%  & $(32\pm11)$\% & $(16\pm 9)$\% & $( 6\pm 4)$\% & $(5 \pm 3)$\% \\
$5\times10^{18}$ &
  $(60\pm14)$\%  & $(21\pm13)$\% & $(10\pm 8)$\% & $( 5\pm 4)$\% & $( 3\pm 3)$\% \\
\botrule
\end{tabular}}
\end{table}

Hence, in the framework of the QGSJET model one can conclude that the mass
composition of PCR changes under the energy range switch-over from
$(5-30)\times10^{17}$~eV to $(5-30)\times10^{18}$~eV. At $\E\ge3\times10^{18}$~eV
the primary cosmic radiation is composed of about $70$\% of protons and helium
nuclei; a fraction of the rest nuclei in the range where there is the second
irregularity in the energetic spectrum (so-called ``ankle'') does not exceed
$\sim30$\%.
A high content of proton and helium nuclei in PCR within the region of
formation of the ``ankle'' is most likely related with a noticeable contribution
of extragalactic cosmic rays into the overall flux of cosmic radiation in
the Earth's vicinity.

\section*{Acknowledgements}

This work has been financially supported by RFBR, grant \#02--02--16380, grant
\#03--02--17160 and grant INTAS \#03--51--5112.

\end{document}